\documentclass[superscriptaddress,twocolumn,preprintnumbers,amsmath,amssymb]{revtex4}
\usepackage[latin1]{inputenc}
\usepackage{amsfonts, amsmath, amsthm}
\usepackage{amssymb, amscd, amstext}
\usepackage{latexsym}
\usepackage[dvips]{graphicx}
\usepackage{subfigure}

\begin{document}

\title{Secure Coherent-state Quantum Key Distribution Protocols with Efficient Reconciliation}
\author{G. \surname{Van Assche}}
\email{gvanassc@ulb.ac.be}
\affiliation{QuIC, Ecole Polytechnique, Université Libre de Bruxelles, CP 165/59, 1050 Brussels, Belgium}
\author{S. \surname{Iblisdir}}
\affiliation{QuIC, Ecole Polytechnique, Université Libre de Bruxelles, CP 165/59, 1050 Brussels, Belgium}
\affiliation{GAP-Optique, University of Geneva, 20 rue de l'Ecole-de-Médecine, CH-1211 Genève, Switzerland}
\author{N. J. \surname{Cerf}}
\affiliation{QuIC, Ecole Polytechnique, Université Libre de Bruxelles, CP 165/59, 1050 Brussels, Belgium}

\begin{abstract}

We study the equivalence between a realistic quantum key distribution protocol using coherent states and homodyne detection and a formal entanglement purification protocol. Maximally-entangled qubit pairs that one can extract in the formal protocol correspond to secret key bits in the realistic protocol. More specifically, we define a qubit encoding scheme that allows the formal protocol to produce more than one entangled qubit pair per coherent state, or equivalently for the realistic protocol, more than one secret key bit. The entanglement parameters are estimated using quantum tomography. We analyze the properties of the encoding scheme and investigate its application to the important case of the attenuation channel.

\end{abstract}

\maketitle
    \section{Introduction}
\label{sectionIntroduction}
    Quantum key distribution (QKD), also called quantum cryptography, allows two parties, Alice and Bob, to share a secret key that can be used for encrypting messages using a classical cipher, e.g., the one-time pad. The main interest of such a key distribution scheme is that any eavesdropping is, in principle, detectable as the laws of quantum mechanics imply that measuring a quantum state generally disturbs it.

The resources required by QKD comprise a source of non-orthogonal quantum states on Alice's side, a quantum channel conveying these states to Bob, a measuring apparatus on Bob's side, and a (public) authenticated classical channel between Alice and Bob. In addition to being used to generate a secret key, the quantum channel is subject to probing by the legitimate parties, so as to determine how many secret bits can be generated.

Most interest in QKD has been devoted to protocols involving (an approximation to) a single-photon source on Alice's side and a single-photon detector on Bob's side (see \cite{bib:entry000} and the references therein). However, protocols involving quantum continuous variables have been considered with an increasing interest \cite{bib:entry001, bib:entry002, bib:entry003, bib:entry004}. Of special importance are coherent-state protocols \cite{bib:entry005, bib:entry006}. The quantum source at Alice's side randomly generates coherent states of a light mode with Gaussian-distributed quadratures, and Bob's measurements are homodyne measurements. These protocols seem to allow for facilitated implementations and higher secret-key generation rates than the protocols involving single-photon sources \cite{bib:entry006}.

Consequently, there is an increasing interest for studying the security of coherent state protocols under general classes of attacks. Individual Gaussian attacks are considered in \cite{bib:entry005, bib:entry006}, and are found to be optimal in the more general class of finite-width non-Gaussian incoherent attacks \cite{bib:entry007}. Individually-probed collective attacks are also considered  in \cite{bib:entry008, bib:entry009}. The recent techniques of \cite{bib:entry00A, bib:entry00B} do not make any assumptions on the eavesdropper's technology and are also considered in \cite{bib:entry008, bib:entry009} for coherent state protocols, although giving lower secret key rates.

In this paper, we study the security of a prepare-and-measure QKD protocol \cite{bib:entry005, bib:entry006} by establishing its equivalence to an entanglement purification (EP) protocol, which produces maximally-entangled qubit pairs. A maximally-entangled qubit pair is by definition completely factored from its environment, and thus the values obtained by measuring each side are fully correlated and secret. The equivalent prepare-and-measure QKD protocol also enjoys this property. This particular technique thus allows one to relieve from any assumptions on the eavesdropper's strategy and was used in \cite{bib:entry00C} to assess the security of the BB84 protocol and in \cite{bib:entry002} for a squeezed state protocol. More recently, this technique was extended to the case of coherent state protocols \cite{bib:entry00D}.

To show the equivalence between a QKD protocol and of an EP protocol, one has to explicitly take into account the secret key distillation, that is, the techniques used to make Alice's and Bob's keys equal (reconciliation) and fully secret (privacy amplification). In \cite{bib:entry00C}, the EP protocol uses CSS quantum codes \cite{bib:entry00E, bib:entry00F}, which are equivalent in QKD to reconciliation with syndromes of binary linear codes and privacy amplification by multiplication with a parity-check matrix. In contrast to BB84, the modulation of coherent states in the protocol we consider here is \emph{continuous}, therefore producing continuous key elements from which to extract a secret key. Reconciliation of a Gaussian-distributed key was studied in \cite{bib:entry010}, and a generic protocol called sliced error correction was designed so as to distill a \emph{binary} key.

In contrast to \cite{bib:entry00D}, the EP protocol is constructed in such a way that it is equivalent to a QKD protocol with sliced error correction for reconciliation. The advantage is the higher secret key rate and the better resistance to attenuation that one can achieve. In particular, more than one maximally-entangled pair (or secret key bit) can be produced per coherent state. Furthermore, thanks to its generality, the asymptotic efficiency of the EP protocol inherits to some extent from the asymptotic efficiency of the classical reconciliation protocol.

The paper is organized as follows. First, in Sec.~\ref{sectionProtocol}, we macroscopically describe the formal EP protocol and its equivalent QKD protocol that are used throughout the paper. Then, in Sec.~\ref{sectionTomography}, we show how the channel can be probed so as to determine the number of secret key bits that Alice and Bob can generate. The encoding of qubits, that is, the generalization of sliced error correction to EP, is described in Sec.~\ref{sectionEncoding}. Then, Sec.~\ref{sectionAttenuationChannel} deals with the important particular case of an attenuation channel. Finally, the asymptotic properties of the qubit encoding scheme are detailed in Sec.~\ref{sectionAsymptotic}.

\section{From Entanglement Purification to Secret Key Distillation}
\label{sectionProtocol}
    After we review the case of EP using CSS codes and its equivalence to BB84, we give a high-level description of a QKD protocol based on EP. We consider this protocol as formal, that is, we do not expect a physical implementation of it. Instead, we propose a prepare-and-measure QKD protocol, derived from the formal one, which also encompasses error correction and privacy amplification.

\subsection{Binary CSS codes}
In the case of BB84, the CSS codes can readily be used to establish the equivalence between an EP protocol and a QKD protocol \cite{bib:entry00C}. Since we will use CSS codes as an ingredient for the EP and QKD protocols below, let us briefly review their properties.

Starting from the Einstein-Podolski-Rosen (EPR) state \begin{equation}|{\phi }^{+}\rangle ={2}^{-1/2}(|00\rangle +|11\rangle )\text{,}\end{equation} Alice keeps half of the state and sends the other half to Bob. His part may undergo a bit error ($|{\phi }^{+}\rangle \to |{\psi }^{+}\rangle $), phase error ($|{\phi }^{+}\rangle \to |{\phi }^{-}\rangle $) or both errors ($|{\phi }^{+}\rangle \to |{\psi }^{-}\rangle $), with $|{\phi }^{-}\rangle ={2}^{-1/2}(|00\rangle -|11\rangle )$ and $|{\psi }^{±}\rangle ={2}^{-1/2}(|01\rangle ±|10\rangle )$. Given that not too many such errors occurs, Alice and Bob can obtain, from many instances of such a transmitted state, a smaller number of EPR pairs using only local operations and classical communications (LOCC). One way to do this is to use CSS codes.

Let ${C}_{1}$ and ${C}_{2}$ be two binary error correcting codes of $n$ bits (i.e., ${C}_{1}$ and ${C}_{2}$ are vector spaces of ${\mathbf{F}}_{2}^{n}$) with parity check matrices ${H}_{1}$ and ${H}_{2}$, resp. They are chosen such that $\{0\}\subset {C}_{2}\subset {C}_{1}\subset {\mathbf{F}}_{2}^{n}$. A CSS code is a $k$-dimensional subspace of ${\mathcal{H}}^{n}$, the Hilbert space of $n$ qubits, with $k=\operatorname{dim}{C}_{1}-\operatorname{dim}{C}_{2}$ \cite{bib:entry00E, bib:entry00F}. The code ${C}_{1}$ allows to correct bit errors, while ${C}_{2}^{\perp }$ (the dual code of ${C}_{2}$) allows to correct phase errors---one important property of the CSS codes is to be able to correct bit errors and phase errors independently.

For entanglement purification, Alice and Bob must compare their syndromes, both for bit errors and phase errors. The relative syndrome determines the correction that Bob must apply to align his qubits to Alice's. Translating this into the BB84 protocol, one can show \cite{bib:entry00C} that the relative syndrome for bit errors in the EP protocol is equal to the relative syndrome for bit errors that Alice and Bob would have reconciled in the BB84 protocol. So, reconciliation can be done using the ${C}_{1}$ code. Phase errors of the EP protocol do not have such a direct equivalent in the BB84 protocol: The prepare-and-measure protocol works as if Alice measured her part of the state in the $\{|0\rangle ,|1\rangle \}$ basis, thereby discarding information on the phase. However, one does not really need to correct the phase errors in the BB84 protocol. Instead, if ${C}_{2}^{\perp }$ would be able to correct them in the EP protocol, the syndrome of ${C}_{2}$ in ${C}_{1}$ of Alice and Bob's bit string turns out to be a valid secret key in the prepare-and-measure protocol. Stated otherwise, ${H}_{1}$ determines the syndrome Alice has to send to Bob to perform reconciliation, while ${H}_{2}$ determines the way the final key is computed for privacy amplification.

Overall, the number of secret key bits is thus $k=\operatorname{dim}{C}_{1}-\operatorname{dim}{C}_{2}$, provided that ${C}_{1}$ (resp. ${C}_{2}^{\perp }$) is small enough to correct all the bit (resp. phase) errors. When considering asymptotically large block sizes, the CSS codes can produce \begin{equation}k=rn\to n(1-h({e}^{\text{b}})-h({e}^{\text{p}}))=Rn\end{equation} EPR pairs or secret key bits, with ${e}^{\text{b}}$ (resp. ${e}^{\text{p}}$) the bit (resp. phase) error rate and $h(p)=-p{\operatorname{log}}_{2}p-(1-p){\operatorname{log}}_{2}(1-p)$ \cite{bib:entry00C}.

We conclude this section by noting that the bit error rate ${e}^{\text{b}}$ determines the number of bits revealed by reconciliation (asymptotically $h({e}^{\text{b}})$), whereas the phase error rate ${e}^{\text{p}}$ determines the number of bits discarded by privacy amplification due to eavesdropping (asymptotically $h({e}^{\text{p}})$).

\subsection{Quantum key distribution based on entanglement purification}
In BB84, the modulation of qubits can be transposed as if Alice prepares a $|{\phi }^{+}\rangle $ state and measures her part. In the case of the QKD protocol with Gaussian-modulated coherent states, the formal state that Alice prepares is of course different, as it must reduce to the proper modulation when Alice measures her part. We define the formal state as:

\begin{equation}
\label{eqPsi}|\Psi \rangle =\int \mathrm{d}x\mathrm{d}p\: g(x,p){|x\rangle }_{{\mathsf{a}}_{1}}\otimes {|p\rangle }_{{\mathsf{a}}_{2}}\otimes {|x+ip\rangle }_{\mathsf{b}}\text{,}\end{equation}where $g(x,p)$ denotes a bi-variate Gaussian distribution $g(x,p)=\sqrt{{G}_{1}(x){G}_{2}(p)}$. The kets $|x\rangle $, $|p\rangle $, $|x+ip\rangle $ are shorthand notations for respectively a $\mathbf{x}$-quadrature eigenstate with eigenvalue $x$, a $\mathbf{p}$-quadrature eigenstate with eigenvalue $p$ and a coherent state whose $\mathbf{x}$ mean value equals $x$ and whose $\mathbf{p}$ mean value equals $p$. The subscripts ${\mathsf{a}}_{1}$, ${\mathsf{a}}_{2}$ (resp. $\mathsf{b}$) denote that the system is lying on Alice's side (resp. Bob's side).

The state (\ref{eqPsi}) does not have a direct physical meaning. In particular, the systems ${\mathsf{a}}_{1}$ and ${\mathsf{a}}_{2}$ must be understood as classical pointers, e.g., resulting from the (formal) homodyne measurement of an EPR state as studied in \cite{bib:entry011}.

In the entanglement purification picture, the $\mathsf{b}$ part of the system is sent to Bob (and possibly attacked by Eve) and the $\mathsf{a}$ part stays at Alice's station. If Alice measures $\mathbf{x}$ in ${\mathsf{a}}_{1}$ and $\mathbf{p}$ in ${\mathsf{a}}_{2}$, the state is projected as if Alice sent Bob a coherent state centered on $x+ip$.

Let us now describe the EP protocol, which reduces to the prepare-and-measure QKD protocol described further.

\begin{itemize}

\item Alice creates $l+n$ copies of the state $|\Psi \rangle $, of which she sends the $\mathsf{b}$ part to Bob.

\item Bob acknowledges reception of the states.
\item Out of the $l+n$ states, $n$ will serve for estimation purposes. These states are chosen randomly and uniformly by Alice, who informs Bob about their positions.
\item For the remaining $l$ states, Alice and Bob perform entanglement purification, so as to produce $rl$ ($0\le r\le 1$) states very close to $|{\phi }^{+}\rangle $. Measured in the computational bases, the produced states yield $rl$ secret bits on both Alice's and Bob's sides.

\end{itemize}
The details of the EP procedure, which uses CSS codes as an ingredient, are given in Sec.~\ref{sectionEncoding}, while the estimation is detailed in Sec.~\ref{sectionTomography}.

\subsection{Prepare-and-measure quantum key distribution}
By virtually measuring the $\mathsf{a}$ part of the state $|\Psi \rangle $, the protocol above reduces to the following one.

\begin{itemize}

\item Alice modulates $l+n$ coherent states $|x+ip\rangle $ that she sends to Bob. The choice of the values of $x$ and $p$ follow the distribution ${|g(x,p)|}^{2}={G}_{1}(x){G}_{2}(p)$.
\item Bob acknowledges reception of the states.
\item Out of the $l+n$ states, $n$ will serve for estimation purposes. These states are chosen randomly and uniformly by Alice, who informs Bob about their positions.
\item For the remaining $l$ states, Bob measures $\mathbf{x}$. Alice and Bob perform secret key distillation (reconciliation and privacy amplification), so as to produce $rl$ secret bits.

\end{itemize}
The reconciliation and privacy amplification procedures are based on classical error correcting codes, which derive from the CSS codes used in the formal EP protocol.

\section{Error Rates Estimation using Tomography}
\label{sectionTomography}
    In QKD protocols derived from EP, an important step is to show how one can infer the bit and phase error rates of the samples that compose the key. A fraction of the samples sent by Alice to Bob are sacrificed so as to serve as test samples. By randomly choosing them within the stream of data, they are statistically representative of the whole stream.

In \cite{bib:entry002, bib:entry00C}, one can simply make measurements and directly count the number of bit and phase errors from the results. This is possible since Bob's apparatus can measure both bit and phase values. In \cite{bib:entry00D}, however, it is not possible to measure directly phase errors. Yet some data post-processing can be applied on measurements so as to infer the number of phase errors in the stream of data. In this section, we wish to show that we can extend this to more general (and more efficient) encodings of qubits (in the EP picture) or bits (in the derived QKD protocol).

The encoding of bits will be described in a further section---for the moment, the qubit pair system (i.e., one among possibly several ones), which Alice and Bob will process using CSS codes, is abstractly represented by its Pauli operators acting in ${\mathsf{a}}_{1}$: ${\mathbf{Z}}_{\mathsf{s}}$ (phase flip) and ${\mathbf{X}}_{\mathsf{s}}$ (bit flip), and in $\mathsf{b}$: ${\mathbf{Z}}_{\mathsf{e}}$ and ${\mathbf{X}}_{\mathsf{e}}$. (The subscripts $\mathsf{s}$ and $\mathsf{e}$ stand for slice and estimator, resp., to follow the convention of the following sections.) The bit errors are assumed to be easy to determine, that is, ${\mathbf{Z}}_{\mathsf{s}}$ has a diagonal expansion in ${|x\rangle }_{{\mathsf{a}}_{1}}\langle x|$, and ${\mathbf{Z}}_{\mathsf{e}}$ can directly be determined by a single homodyne measurement on $\mathsf{b}$. This ensures, in the derived prepare-and-measure QKD protocol, that Alice knows the bit value she sent, and Bob can determine the received bit value. A measurement of the observable ${\mathbf{X}}_{\mathsf{s}}{\mathbf{I}}_{{\mathsf{a}}_{2}}{\mathbf{X}}_{\mathsf{e}}$ associated to the phase error rate, however, cannot be implemented by a single homodyne measurement on $\mathsf{b}$. Therefore, we have to invoke quantum tomography with a quorum of operators \cite{bib:entry012} to get an estimate of the phase error rate.

\subsection{Estimating phase errors in the average state}
In the EP picture, let ${\rho }^{(n)}$ be the state of the $n$ samples used for estimation of the phase error rate (i.e., $n$ instances of the ${\mathsf{a}}_{1}{\mathsf{a}}_{2}\mathsf{b}$ system). To count the number of phase errors in a set of $n$ samples, one needs to measure $\mathbf{O}={\mathbf{X}}_{\mathsf{s}}{\mathbf{I}}_{{\mathsf{a}}_{2}}{\mathbf{X}}_{\mathsf{e}}$ on the $n$ samples and sum the results (with ${\mathbf{I}}_{{\mathsf{a}}_{2}}$ the identity in the system ${\mathsf{a}}_{2}$). This is equivalent to measuring ${\mathbf{O}}^{(n)}={\sum }_{i}{\mathbf{I}}_{{\mathsf{a}}_{1}{\mathsf{a}}_{2}\mathsf{b}}^{\otimes i-1}\otimes {\mathbf{X}}_{\mathsf{s}}{\mathbf{I}}_{{\mathsf{a}}_{2}}{\mathbf{X}}_{\mathsf{e}}\otimes {\mathbf{I}}_{{\mathsf{a}}_{1}{\mathsf{a}}_{2}\mathsf{b}}^{\otimes n-i}$. If the true phase error probability in the $n+l$ samples is ${e}^{\text{p}}$, the error variance is ${\sigma }_{1}^{2}=2{e}^{\text{p}}(1-{e}^{\text{p}})/n$, and thus the probability of making an estimation error of more than $\Delta $ is \cite{bib:entry002, bib:entry00C} asymptotically $\operatorname{exp}(-{\Delta }^{2}n/4{e}^{\text{p}}(1-{e}^{\text{p}}))$. It is easy to see that \begin{equation}\operatorname{Tr}({\mathbf{O}}^{(n)}{\rho }^{(n)})=n\operatorname{Tr}(\mathbf{O}\rho )\text{,}\end{equation} where $\rho ={n}^{-1}{\sum }_{i}{\operatorname{Tr}}_{\text{All}\setminus \{i\}}({\rho }^{(n)})$ is the density matrix of the average state measured. So, we can estimate the number of phase errors using the average state, even if eavesdropping is joint (${\rho }^{\otimes n}\ne {\rho }^{(n)}$).

If the measurement of $\mathbf{O}={\mathbf{X}}_{\mathsf{s}}{\mathbf{I}}_{{\mathsf{a}}_{2}}{\mathbf{X}}_{\mathsf{e}}$ cannot be made directly, one instead looks for a quorum of operators ${\mathbf{Q}}_{\lambda }$ such that $\mathbf{O}=\int \mathrm{d}\lambda o(\lambda ){\mathbf{Q}}_{\lambda }$; estimating $\langle \mathbf{O}\rangle $ comes down to measuring several times ${\mathbf{Q}}_{\lambda }$ for values of $\lambda $ chosen randomly and independently of each other, and averaging the results weighted by $o(\lambda )$: $\mathbf{O}\approx {\sum }_{i}o({\lambda }_{i}){\mathbf{Q}}_{{\lambda }_{i}}$ \cite{bib:entry012}. If the values of $\lambda $ are chosen independently of the sample index on which ${\mathbf{Q}}_{\lambda }$ is applied, we get unbiased results, as $\operatorname{Tr}(\mathbf{O}\rho )={E}_{\lambda }[\operatorname{Tr}({\mathbf{Q}}_{\lambda }\rho )]$, with $E$ the expectation. Of course, the estimation of $\operatorname{Tr}(\mathbf{O}\rho )$ with a quorum cannot be perfect and an estimation variance ${\sigma }_{2}^{2}$ must also be considered and added, ${\sigma }^{2}={\sigma }_{1}^{2}+{\sigma }_{2}^{2}$.

\subsection{Estimating phase errors using coherent states and homodyne detection}
We now explain how the phase error rate can be estimated, in principle, using coherent states modulated in both quadratures and homodyne detection in all quadratures.

 It is clear that the knowledge of matrix elements of the average state $\rho $ gives the knowledge of $\langle \mathbf{O}\rangle $. Let ${\rho }_{0}=|\Psi \rangle \langle \Psi |$ be the state that Alice and Bob would share if the transmission was perfect. Since the $\mathsf{a}$ part of the system stays at Alice's station, we only need to learn about how the $\mathsf{b}$ part of the system is affected. In the prepare-and-measure picture, let $T$ be the completely positive (CP) map that maps the states sent by Alice onto the states received by Bob, $(\operatorname{Id}\otimes T)({\rho }_{0})=\rho $. In particular, let the coherent state $|x+ip\rangle \langle x+ip|$ be mapped onto ${\rho }_{T}(x+ip)$ and the (pseudo-)position state $|x\rangle \langle {x}^{\prime }|$ be mapped onto ${\rho }_{T}(x,{x}^{\prime })$. The functions ${\rho }_{T}(x+ip)$ and ${\rho }_{T}(x,{x}^{\prime })$ are related by the following identity: \begin{multline}{\rho }_{T}(x+ip)\propto \int \mathrm{d}{x}^{\prime }\mathrm{d}{x}^{\prime \prime }{e}^{-{({x}^{\prime }-x)}^{2}/4{N}_{0}-{({x}^{\prime \prime }-x)}^{2}/4{N}_{0}}\\
{e}^{i({x}^{\prime }-{x}^{\prime \prime })p/2{N}_{0}}{\rho }_{T}({x}^{\prime },{x}^{\prime \prime })\text{,}\end{multline} with ${N}_{0}$ the variance of the vacuum fluctuations. By setting $D={x}^{\prime }-{x}^{\prime \prime }$ and $S={x}^{\prime }+{x}^{\prime \prime }-2x$, we get: \begin{multline}
\label{eqrhoTSD}{\rho }_{T}(x+ip)\propto \int \mathrm{d}D\mathrm{d}S{e}^{-{S}^{2}/8{N}_{0}-{D}^{2}/8{N}_{0}+iDp/2{N}_{0}}\\
{\rho }_{T}(x+S+D,x+S-D)\text{,}\end{multline} which shows that ${\rho }_{T}(x,{x}^{\prime })$ is integrated with an invertible kernel (Gaussian convolution in $S$, multiplication by ${e}^{-{D}^{2}/8{N}_{0}}$, and Fourier transform in $D$). So in principle, any different CP map ${T}^{\prime }\ne T$ implies a different effect on coherent states, ${\rho }_{T}(x+ip)\ne {\rho }_{{T}^{\prime }}(x+ip)$. The modulation of coherent states in both quadratures is thus crucial for this implication being possible.

By inspecting Eq.~(\ref{eqrhoTSD}), it seems that due to the factors ${e}^{-{S}^{2}/8{N}_{0}}$ and ${e}^{-{D}^{2}/8{N}_{0}}$, two different CP-maps $T$ and ${T}^{\prime }$ may make ${\rho }_{T}(x+ip)$ and ${\rho }_{{T}^{\prime }}(x+ip)$ only vanishingly different. It thus seems unlikely that Eq.~(\ref{eqrhoTSD}) should allow us to extract the coefficients ${\rho }_{T}(x+S+D,x+S-D)$. However, assuming that $T$ depends only on a finite number of parameters, a variation of these parameters will induce a measurable variation of ${\rho }_{T}(x+ip)$. We will now discuss why it is reasonable to make such an assumption.

Due to the finite variance of the modulation of coherent states, the probability of emission of a large number of photons vanishes---this intuitively indicates that we only need to consider the description of $T$ for a bounded number of emitted photons. More precisely, one can consider the emission of $w$ joint copies of the state ${\rho }_{0\mathsf{b}}={\operatorname{Tr}}_{\mathsf{a}}({\rho }_{0})$. For $w$ sufficiently large ${\rho }_{0\mathsf{b}}^{\otimes w}$ can be represented in the typical subspace ${\Gamma }_{\delta }({\rho }_{0\mathsf{b}})$ of dimension not greater than ${2}^{w(H({\rho }_{0\mathsf{b}})+\delta )}$, for any $\delta >0$ \cite{bib:entry013}, where $H(\rho )$ is the Von Neumann entropy of a state $\rho $. The probability mass of ${\rho }_{0\mathsf{b}}^{\otimes w}$ outside the typical subspace can be made arbitrarily small and does not depend on the eavesdropping strategy. This means that the support for the input of $T$ has finite dimension, up to an arbitrarily small deviation.

The number of photons received by Bob can also be upper bounded. Alice and Bob can first assume that no more than ${n}_{\text{max}}$ photons are received. This fact may depend on a malicious eavesdropper, so Bob has to do hypothesis testing. The test comes down to estimating $\langle \Pi \rangle $ with $\Pi ={\sum }_{n>{n}_{\text{max}}}|n\rangle \langle n|$. If the threshold is well chosen so that $n>{n}_{\text{max}}$ never occurs, we can apply the central limit theorem and upper bound the probability that $\langle \Pi \rangle >\epsilon $ for any chosen $\epsilon >0$. The positivity of the density matrices implies that the off-diagonal coefficients are also bounded. We can thus now express ${\rho }_{T}(x+ip)$ as ${\rho }_{T}(x+ip)={\sum }_{n,{n}^{\prime }\le {n}_{\text{max}}}{\rho }_{T}(x+ip,n,{n}^{\prime })|n\rangle \langle {n}^{\prime }|$. Note that the test can be implemented either by explicitly measuring the intensity of the beam (therefore requiring an additional photodetector) or by exploiting the correlation between the high intensity of the beam and the high absolute values obtained when doing homodyne measurements in all directions.

Finally, the estimation of the coefficient of $|n\rangle \langle {n}^{\prime }|$ can be done with arbitrarily small statistical error using homodyne detection in all directions \cite{bib:entry012, bib:entry014}. This is achieved by considering the quorum of operators ${({\mathbf{x}}_{\theta })}_{0\le \theta <2\pi }$, where ${\mathbf{x}}_{\theta }=\operatorname{cos}\theta \: \mathbf{x}+\operatorname{sin}\theta \: \mathbf{p}$ denotes the amplitude of the quadrature in direction $\theta $. Considering a finite combination of arbitrarily small statistical errors on parameters also gives arbitrarily small overall statistical error on the phase error rate.

\section{Encoding of Multiple QuBits in an Oscillator}
\label{sectionEncoding}
    Reconciliation and privacy amplification are integral parts of the prepare-and-measure protocols derived from entanglement purification protocols. In our case, we wish to derive a prepare-and-measure protocol with sliced error correction (SEC) \cite{bib:entry010} as reconciliation, which allows us to obtain a higher secret key rate and a better resistance to losses than in \cite{bib:entry00D}. We therefore need to describe an entanglement purification procedure that reduces to SEC when the corresponding prepare-and-measure protocol is derived. An overview of SEC is proposed next.

\subsection{Sliced error correction with invertible mappings}
We here recall the main principles of SEC in a form that is slightly different from the presentation in \cite{bib:entry010}. To suit our needs, we here describe SEC in terms of invertible functions giving the slices and the estimators---the invertibility property will be required when we generalize SEC to entanglement purification. Also, from the generality of \cite{bib:entry010}, two parameters are fixed here: Binary error correction is operated by sending syndromes of classical linear error-correcting codes (ECC), and we momentarily restrict ourselves to the case of scalar values.

Suppose Alice and Bob have $l$ pairs of correlated random variables $({X}_{1},{X}_{1}^{\prime }),\dots ,({X}_{l},{X}_{l}^{\prime })$, with ${X}_{i},{X}_{i}^{\prime }\in \mathbf{R}$, $i=1\dots l$, from which they intend to extract common bits.

First, Alice wishes to convert each of her variables $X$ into $m$ bits and thereby defines $m$ binary functions: ${S}_{1}(X),\: \dots ,{S}_{m}(X)$. To make the mapping invertible, she also defines a function $\bar{S}(X)$ such that mapping from $X$ to the vector $(\bar{S}(X),{S}_{1\dots m}(X))$ is bijective. As a convention, the range of $\bar{S}(X)$ is $[0;1]$. The mapping from $\mathbf{R}$ to $[0;1]×{\{0,1\}}^{m}$: \begin{equation}x\to (\bar{S}(x),{S}_{1\dots m}(x))\end{equation} is collectively denoted as $\mathcal{S}$.

Concretely, the functions ${S}_{i}(X)$ implicitly cut the real line into intervals (see \cite{bib:entry010} for more details), whereas $\bar{S}(X)$ indicates where to find $X$ within a given interval.

Then, we can assemble the bits produced by the $l$ random variables ${X}_{1}\dots {X}_{l}$ into $m$ $l$-bit vectors. To each bit vector ("slice") ${S}_{i}({X}_{1\dots l})=({S}_{i}({X}_{1}),\dots ,{S}_{i}({X}_{l}))$, is associated an ECC that Alice and Bob agreed upon. To proceed with the correction, Alice sends the syndrome ${\xi }_{i}^{\text{b}}={H}_{i}^{\text{b}}{S}_{i}({X}_{1\dots l})$ to Bob over the public authenticated channel, where ${H}_{i}^{\text{b}}$ is the ${l}_{i}^{\text{b}}×l$ parity check matrix of the ECC associated to slice $i$. Alice also sends $\bar{S}({X}_{1\dots l})$.

 Bob would like to recover ${S}_{1\dots m}({X}_{1\dots l})$ from his knowledge of ${X}_{1\dots l}^{\prime }$, ${\xi }_{1\dots m}^{\text{b}}$ and $\bar{S}({X}_{1\dots l})$. To do so, he also converts each of his variables ${X}_{1\dots l}^{\prime }$ into $m$ bits, but he does so in a consecutive manner. He tries to produce bits that are best correlated to Alice's and takes advantage of the corrected bits of slices $j<i$ before trying to estimate the bits of slice $i$. In particular, to produce bits that are best correlated to Alice's first slice ${S}_{1}({X}_{1\dots l})$, he uses a function ${E}_{1}({X}^{\prime },\bar{S}(X))$, which gives his best estimate on Alice's corresponding bit ${S}_{1}(X)$ given the known correlations between $X$ and ${X}^{\prime }$. By applying the function ${E}_{1}$ on all the variables ${X}_{1\dots l}^{\prime }$ and $\bar{S}({X}_{1\dots l})$, Bob is able to construct a string of $l$ bits that is equal to Alice's up to some error rate ${e}_{1}^{\text{b}}$. Given the knowledge of ${\xi }_{1}^{\text{b}}$ and assuming the adequacy of the ECC, Bob has enough information to determine ${S}_{1}({X}_{1\dots l})$ with high probability. Then, for slice $i>1$, he estimates ${S}_{i}({X}_{1\dots l})$ using the estimator function ${E}_{i}({X}^{\prime },\bar{S}(X),{\beta }_{1},\dots ,{\beta }_{i-1})$, where ${\beta }_{j}$ is the random variable indicating Bob's knowledge of ${S}_{j}(X)$, so that ${\beta }_{j}={S}_{j}(X)$ with arbitrarily high probability. (Note that the estimators can also be written as jointly working on $l$ samples at once: ${E}_{i}({X}_{1\dots l}^{\prime },\bar{S}({X}_{1\dots l}),{\xi }_{1}^{\text{b}},\dots ,{\xi }_{i-1}^{\text{b}})$, but we will preferably use the previous notation for its simplicity since, besides the ECC decoding, all the operations are done on each variable $X$ or ${X}^{\prime }$ independently.)

 We also need a supplementary function to ensure that the process on Bob's side is described using bijective functions: $\bar{E}({X}^{\prime },\bar{S}(X),{\beta }_{1},\dots ,{\beta }_{m})$ (or jointly $\bar{E}({X}_{1\dots l}^{\prime },\bar{S}({X}_{1\dots l}),{\xi }_{1}^{\text{b}},\dots ,{\xi }_{m}^{\text{b}})$). As a convention, the range of $\bar{E}$ is $[0;1]$. $\bar{E}$ is chosen so that the mapping $\mathcal{E}$ defined below is invertible, \begin{multline}\mathcal{E}\text{:}(\bar{s},{s}_{1\dots m}^{\prime },{x}^{\prime })\to (\bar{s},{s}_{1\dots m}^{\prime },{E}_{1}({x}^{\prime },\bar{s}),\dots ,\\
{E}_{m}({x}^{\prime },\bar{s},{s}_{1\dots m-1}^{\prime }),\bar{E}({x}^{\prime },\bar{s},{s}_{1\dots m}^{\prime }))\text{.}\end{multline}

Similarly to $\mathcal{S}$, the functions ${E}_{1\dots m}$ of $\mathcal{E}$ cut the real line into intervals. However, these intervals are adapted as a function of the information sent by Alice, so as to estimate Alice's bits more reliably. Like for $\bar{S}$, the function $\bar{E}$ indicates where to find ${X}^{\prime }$ within an interval.

 The mapping $\mathcal{S}$ summarizes Alice's process of conversion of her real values $X$ into $m$ bits (plus a continuous component). The mapping $\mathcal{E}$ represents the bits (and a continuous component) produced by Bob from his real values ${X}^{\prime }$ and his knowledge of $\bar{S}(X)$ and of the syndromes ${\xi }_{1\dots m}^{\text{b}}$. The bits produced by the functions ${E}_{i}$ are not yet corrected by the ECC, even though they take as input the corrected values of the previous slices ${S}_{j}(X)$, $j<i$. The description of the mapping $\mathcal{E}$ with the bits prior to ECC correction allows us to easily express the bit error rate between Alice's slices and Bob's estimators and thereby to deduce the size of the parity matrices of the ECCs needed for the binary correction. Simply, we define ${e}_{i}^{\text{b}}=\operatorname{Pr}[{S}_{i}(X)\ne {E}_{i}({X}^{\prime },\bar{S}(X),{S}_{1\dots i-1}(X))]$. As the block size $l\to \infty $, there exist ECCs with size ${l}_{i}^{\text{b}}\to lh({e}_{i}^{\text{b}})$ and arbitrarily low probability of decoding error. The number of common (but not necessarily secret) bits produced by SEC is therefore asymptotically equal to $H({S}_{1\dots m}(X))-{\sum }_{i=1}^{m}h({e}_{i}^{\text{b}})$ per sample \cite{bib:entry010}.

The generalization of the SEC to a quantum entanglement purification protocol is examined next.

\subsection{Quantum sliced error correction}
From classical binary error correcting codes, one can construct CSS quantum codes and use them to extract EPR pairs from noisy qubit pairs. We will now show that, similarly, from SEC, it is possible to construct an encoding and decoding procedure, which when applied on entangled quantum oscillator systems, also allows to extract pure EPR pairs. Such a purification protocol is formal, as it would of course be very difficult to implement in practice.

The purification uses a few quantum registers, which we now list. Alice's system ${\mathsf{a}}_{1}$ is split into $m$ qubit systems ${\mathsf{s}}_{1\dots m}$ and a continuous register $\bar{\mathsf{s}}$. On Bob's side, the system $\mathsf{b}$ is split into $m$ qubit systems ${\mathsf{e}}_{1\dots m}$ and a continuous register $\bar{\mathsf{e}}$. He also needs $m$ qubit registers ${\mathsf{s}}_{1\dots m}^{\prime }$ for temporary storage. All these registers must of course be understood per exchanged sample: As Alice generates $l$ copies of the state $|\Psi \rangle $, the legitimate parties use $l$ instances of the registers listed above.

The usual bit-flip and phase-flip operators $\mathbf{X}$ and $\mathbf{Z}$, resp., can be defined as acting on a specific qubit register among the systems ${\mathsf{s}}_{i}$ and ${\mathsf{e}}_{i}$. E.g., ${\mathbf{Z}}_{{\mathsf{s}}_{i}}$ is defined as acting on ${\mathsf{s}}_{i}$ only. These operators are used by Alice and Bob to construct the CSS codes that produce entangled qubits, which are in turn used to produce EPR pairs in the registers ${\mathsf{s}}_{i}{\mathsf{e}}_{i}$ for $i=1\dots m$. Since each CSS code operates in its own register pair, the action of one does not interfere with the action of the other. It is thus possible to extract more than one EPR pair $|{\phi }^{+}\rangle $ per state $|\Psi \rangle $. If asymptotically efficient binary codes are used, the rate or EPR pairs produced is $R={\sum }_{i}(1-h({e}_{i}^{\text{b}})-h({e}_{i}^{\text{p}}))$, where ${e}_{i}^{\text{b}}$ (resp. ${e}_{i}^{\text{p}}$) indicates the bit error rate (resp. the phase error rate) \cite{bib:entry00C}.

The process that defines the content of the registers is described next.

\subsubsection{The mappings $\mathcal{QS}$ and $\mathcal{QE}$}
\begin{figure}
\begin{center}

\includegraphics[width=4cm]{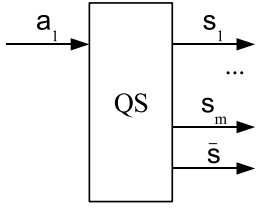}

\end{center}

\caption{Schematic description of $\mathcal{QS}$}
\label{figureQS}
\end{figure}

First, we define the unitary transformation $\mathcal{QS}$: ${L}^{2}(\mathbf{R})\to {L}^{2}([0;1])\otimes {\mathcal{H}}^{\otimes m}$ by its application on the basis of quadrature eigenstates: \begin{equation}
\label{eq_QS}{|x\rangle }_{{\mathsf{a}}_{1}}\to \sigma (x){|\bar{S}(x)\rangle }_{\bar{\mathsf{s}}}\otimes {|{S}_{1}(x)\rangle }_{{\mathsf{s}}_{1}}\otimes \dots \otimes {|{S}_{m}(x)\rangle }_{{\mathsf{s}}_{m}}\text{.}\end{equation} The states ${|\bar{s}\rangle }_{\bar{\mathsf{s}}}$, $0\le \bar{s}\le 1$, form an orthogonal basis of ${L}^{2}([0;1])$, $\sigma (x)={({\mathrm{d}}_{x}\bar{S})}^{-1/2}(x)$ is a normalization function, and ${|{s}_{i}\rangle }_{{\mathsf{s}}_{i}}$, ${s}_{i}\in \{0,1\}$, denotes the canonical basis of $\mathcal{H}$, the Hilbert space of a qubit. As a convention, the system ${\mathsf{s}}_{i}$ is called slice~$i$. The transformation $\mathcal{QS}$ is depicted in Fig.~\ref{figureQS}.

 For each slice $i$, Alice and Bob agree on a CSS code, defined by its parity matrices ${H}_{i}^{\text{b}}$ for bit error correction and ${H}_{i}^{\text{p}}$ for phase error correction. For the entanglement purification, let us assume that Alice computes the syndromes of the CSS code with a quantum circuit. For each slice, she produces ${l}_{i}^{\text{b}}$ qubits in the state $|{\xi }_{i}^{\text{b}}\rangle $ and ${l}_{i}^{\text{p}}$ qubits in the state $|{\xi }_{i}^{\text{p}}\rangle $ that she sends to Bob over a perfect quantum channel, so that the syndromes are received without any distortion. In the entanglement purification picture, the syndromes can be transmitted over a non-perfect channel if they are encoded using appropriate error correcting codes. Also, after reduction to a prepare-and-measure protocol, this perfect transmission is actually done over the public authenticated channel. Alice also sends the $\bar{\mathsf{s}}$ system to Bob.

 Then, the slice estimators are defined as the unitary transformation $\mathcal{QE}$ from ${L}^{2}([0;1])\otimes {\mathcal{H}}^{\otimes m}\otimes {L}^{2}(\mathbf{R})$ to ${L}^{2}([0;1])\otimes {\mathcal{H}}^{\otimes m}\otimes {\mathcal{H}}^{\otimes m}\otimes {L}^{2}([0;1])$: \begin{multline}
\label{eq_QE}{|\bar{s}\rangle }_{\bar{\mathsf{s}}}{|{s}_{1\dots m}^{\prime }\rangle }_{{s}_{1\dots m}^{\prime }}{|{x}^{\prime }\rangle }_{\mathsf{b}}\to \epsilon ({x}^{\prime },\bar{s},{s}_{1\dots m}^{\prime }){|\bar{s}\rangle }_{\bar{\mathsf{s}}}{|{s}_{1\dots m}^{\prime }\rangle }_{{\mathsf{s}}_{1\dots m}^{\prime }}\\
{\otimes }_{i=1}^{m}{|{E}_{i}({x}^{\prime },\bar{s},{s}_{1\dots i-1}^{\prime })\rangle }_{{\mathsf{e}}_{i}}{|\bar{E}({x}^{\prime },\bar{s},{s}_{1\dots m}^{\prime })\rangle }_{\bar{\mathsf{e}}}\text{,}\end{multline} where $\epsilon ({x}^{\prime },\bar{s},{s}_{1\dots m}^{\prime })={({\partial }_{{x}^{\prime }}\bar{E})}^{-1/2}({x}^{\prime },\bar{s},{s}_{1\dots m}^{\prime })$ is a normalization function; ${|{x}^{\prime }\rangle }_{\mathsf{b}}$ is a quadrature eigenstate with $\mathbf{x}$-eigenvalue ${x}^{\prime }$; ${|{e}_{i}\rangle }_{{\mathsf{e}}_{i}}$, ${e}_{i}\in \{0,1\}$, denotes the canonical basis of $\mathcal{H}$; ${|\bar{e}\rangle }_{\bar{\mathsf{e}}}$, $0\le \bar{e}\le 1$, form an orthogonal basis of ${L}^{2}([0;1])$. As the classical mapping $\mathcal{E}$ is invertible, $\mathcal{QE}$ is unitary with the appropriate normalization function $\epsilon $. This mapping is defined to act on individual states, with the slice values ${s}_{1\dots m}^{\prime }$ as input in the system ${\mathsf{s}}_{1\dots m}^{\prime }$, whose purpose is actually to hold Bob's sequentially corrected bit values ${\beta }_{1\dots m}$. The complete transformation jointly involving $l$ systems would be fairly heavy to describe. Only the ECC correction needs to be described jointly, and assuming it is correctly sized (i.e., ${l}_{i}^{\text{b}}$ are large enough), Bob has enough information to reconstruct Alice's bit values. Let us now sketch how the system ${\mathsf{s}}_{1\dots m}^{\prime }$ is constructed.

\begin{figure}
\begin{center}

\includegraphics[width=8cm]{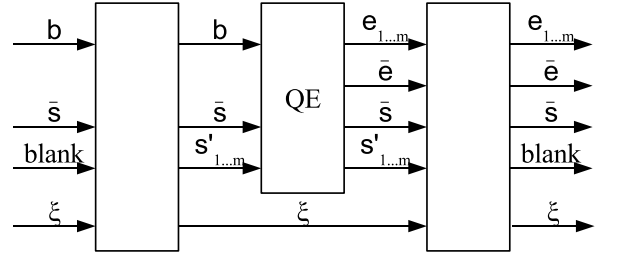}

\end{center}

\caption{Schematic description of $\mathcal{QE}$ and the use of the systems ${\mathsf{s}}_{1\dots m}^{\prime }$}
\label{figureQE}
\end{figure}
 Assume that Bob first calculates, using a quantum circuit, the first slice estimator (classically: ${E}_{1}({X}^{\prime },\bar{S}(X))$), which does not depend on any syndrome. That is, he applies the following mapping, defined on the bases of $\bar{\mathsf{s}}$ and $\mathsf{b}$: ${|\bar{s}\rangle }_{\bar{\mathsf{s}}}{|{x}^{\prime }\rangle }_{\mathsf{b}}\to {|\bar{s}\rangle }_{\bar{\mathsf{s}}}{|{E}_{1}({x}^{\prime },\bar{s})\rangle }_{{\mathsf{e}}_{1}}{|{\bar{E}}_{1}({x}^{\prime },\bar{s})\rangle }_{{\bar{\mathsf{e}}}_{1}}$ (up to normalization), where the function ${\bar{E}}_{1}$ is needed only to make the mapping unitary. From the $l$ qubits in the $l$ systems ${\mathsf{e}}_{1}$ and the syndrome sent by Alice $|{\xi }_{1}^{\text{b}}\rangle $, there exists a quantum circuit that calculates the relative syndrome of Alice's and Bob's bits, that is a superposition of the classical quantities ${\xi }_{1}^{\text{b}}\oplus {H}_{1}^{\text{b}}{E}_{1}({X}_{1\dots l})$. From this, a quantum circuit calculates the coset leader of the syndrome, that is (a superposition of) the most probable difference vector between Alice's and Bob's qubits. An extra $l-{l}_{1}^{\text{b}}$ blank qubits are needed for this operations; we assume they are all initialized to $|0\rangle $: \begin{equation}{|{H}_{1}^{\text{b}}({s}_{1}^{(l)}\oplus {e}_{1}^{(l)})\rangle }_{{\mathsf{s}}_{1}^{\prime ({l}_{1}^{\text{b}})}}{|0\rangle }_{{\mathsf{s}}_{1}^{\prime (l-{l}_{1}^{\text{b}})}}\to {|{s}_{1}^{(l)}\oplus {e}_{1}^{(l)}\rangle }_{{\mathsf{s}}_{1}^{\prime (l)}}\text{.}\end{equation} Then, using a controlled-not operation between Bob's bits (control) and the difference vector (target), we produce $l$ qubits containing the same bit values as Alice's, with an arbitrarily large probability: \begin{equation}{|{e}_{1}^{(l)}\rangle }_{{\mathsf{e}}_{1}^{(l)}}{|{s}_{1}^{(l)}\oplus {e}_{1}^{(l)}\rangle }_{{\mathsf{s}}_{1}^{\prime (l)}}\to {|{e}_{1}^{(l)}\rangle }_{{\mathsf{e}}_{1}^{(l)}}{|{s}_{1}^{(l)}\rangle }_{{\mathsf{s}}_{1}^{\prime (l)}}\text{.}\end{equation} This is how the $l$ systems ${\mathsf{s}}_{1}^{\prime }$ are created.

Following this approach for the next slices, we can define: ${|\bar{s}\rangle }_{\bar{\mathsf{s}}}{|{s}_{1}\rangle }_{{\mathsf{s}}_{1}^{\prime }}{|{E}_{1}({x}^{\prime },\bar{s})\rangle }_{{\mathsf{e}}_{1}}{|{\bar{E}}_{1}({x}^{\prime },\bar{s})\rangle }_{{\bar{\mathsf{e}}}_{1}}\to {|\bar{s}\rangle }_{\bar{\mathsf{s}}}{|{s}_{1}\rangle }_{{\mathsf{s}}_{1}^{\prime }}{|{E}_{1}({x}^{\prime },\bar{s})\rangle }_{{\mathsf{e}}_{1}}{|{E}_{2}({x}^{\prime },\bar{s},{s}_{1})}_{{\mathsf{e}}_{2}}{|{\bar{E}}_{2}({x}^{\prime },\bar{s},{s}_{1})\rangle }_{{\bar{\mathsf{e}}}_{2}}$, and reasonably assume that the bit value given in ${\mathsf{s}}_{1}^{\prime }$ is equal to Alice's ${S}_{1}(X)$. This reasoning can be applied iteratively, so as to fill the system ${\mathsf{s}}_{1\dots m}^{\prime }$ with all the corrected bit values, and with an extra step to set $\bar{E}({x}^{\prime },\bar{s},{s}_{1\dots m})$ in $\bar{\mathsf{e}}$.

As a last step, Bob can revert the ECC decoding operations and come back to the situation where he has blank qubits in ${\mathsf{s}}_{1\dots m}^{\prime }$ as depicted in Fig.~\ref{figureQE}.

Finally, the qubits produced by $\mathcal{QE}$ can be transformed into EPR pairs using the CSS codes and the syndromes Alice sent to Bob.

\subsubsection{Phase coherence}
\label{sec_phase_coherence}
    Neither the unitary transformation $\mathcal{QS}$ nor $\mathcal{QE}$ take into account the modulation of the coherent state in the $\mathbf{p}$-quadrature. By ignoring what happens in the ${\mathsf{a}}_{2}$ system of Eq.~(\ref{eqPsi}), the reduced system ${\rho }_{{\mathsf{a}}_{1}\mathsf{b}}$ lacks phase coherence: \begin{multline}{\rho }_{{\mathsf{a}}_{1}\mathsf{b}}=\int \mathrm{d}x\mathrm{d}{x}^{\prime }\mathrm{d}p\sqrt{{G}_{1}(x){G}_{1}({x}^{\prime })}{G}_{2}(p){|x\rangle }_{{\mathsf{a}}_{1}}\langle {x}^{\prime }|\\
D(ip){|x+i0\rangle }_{\mathsf{b}}\langle {x}^{\prime }+i0|{D}^{\dagger}(ip)\text{.}\end{multline} To remedy this, we assume that Alice also sends the ${\mathsf{a}}_{2}$ system to Bob, just like she does for the $\bar{\mathsf{s}}$ system and the syndromes, since the modulation in the $\mathbf{p}$-quadrature is independent of the key. Bob can take it into account before applying $\mathcal{QE}$, by displacing his state along the $\mathbf{p}$-quadrature in order to bring it on the $\mathbf{x}$-axis.

Actually, we could formally include this ${\mathsf{a}}_{2}$-dependent operation in the $\mathcal{QE}$ mapping, by adding ${|p\rangle }_{{\mathsf{a}}_{2}}$ to its input and output (unmodified) and by multiplying by a factor of the form ${e}^{i{x}^{\prime }p/4{N}_{0}}$ in Eq.~(\ref{eq_QE}), with ${N}_{0}$ the vacuum fluctuations. For notation simplicity, however, we mention it here without explicitly writing it.

Also, for the simplicity of the notation in the next section, we can assume without loss of generality that the coefficients of $|\Psi \rangle $ in the $\mathbf{x}$-basis of $\mathsf{b}$ are real, after adjustment by Bob as a function of $p$.

\subsubsection{Construction of $\bar{S}$ and $\bar{E}$}
\label{sectionConstructionSbarAndEbar}
    Let us now make explicit the construction of the functions $\bar{S}$ and $\bar{E}$. First assume, for simplicity, that we have only one slice ($m=1$)---for this we do not write the slice index as a subscript. The mapping has thus the following form: \begin{multline}{|x\rangle }_{{\mathsf{a}}_{1}}{|{x}^{\prime }\rangle }_{\mathsf{b}}\to \sigma (x){|S(x)\rangle }_{\mathsf{s}}{|\bar{S}(x)\rangle }_{\bar{\mathsf{s}}}\epsilon ({x}^{\prime },\bar{S}(x),S(x))\\
{|E({x}^{\prime },\bar{S}(x))\rangle }_{\mathsf{e}}{|\bar{E}({x}^{\prime },\bar{S}(x),S(x))\rangle }_{\bar{\mathsf{e}}}\text{,}\end{multline} where $\sigma (x)={({\mathrm{d}}_{x}\bar{S})}^{-1/2}(x)$, $\epsilon ({x}^{\prime },\bar{s},s)={({\partial }_{{x}^{\prime }}\bar{E})}^{-1/2}({x}^{\prime },\bar{s},s)$, and $\bar{S}$ and $\bar{E}$ range between $0$ and $1$.

Let us take some state $\rho $ of the systems $\mathsf{s}\bar{\mathsf{s}}\mathsf{e}\bar{\mathsf{e}}$. In the entanglement purification picture, our goal is to be able to extract entangled pairs in the subsystem ${\rho }_{\mathsf{s}\mathsf{e}}={\operatorname{Tr}}_{\text{All}\setminus \{\mathsf{s},\mathsf{e}\}}(\rho )$. We thus want $\rho $ to be a product state of the form ${\rho }_{\mathsf{s}\mathsf{e}}\otimes {\rho }_{\bar{\mathsf{s}}\bar{\mathsf{e}}}$. If $\bar{S}(X)$ contains information about $S(X)$, or if $\bar{E}({X}^{\prime },\bar{S}(X),S(X))$ contains information about $E({X}^{\prime },\bar{S}(X))$, the subsystem ${\rho }_{\mathsf{s}\mathsf{e}}$ will not be pure. In the prepare-and-measure picture, information on $S(X)$ in $\bar{S}(X)$ will be known to Eve and therefore may not be considered as secure. Note that information in $\bar{E}(\dots )$ is not disclosed, but since it is excluded from the subsystems from which we wish to extract entanglement (or secrecy), any correlation with $\bar{\mathsf{e}}$ will reduce the number of entangled qubits (or secret bits); or stated otherwise, the calculated number of secret bits will be done as if $\bar{E}(\dots )$ was public. As an extreme example, if $S(X)$ and $E({X}^{\prime },\bar{S}(X))$ are perfectly correlated and if $S(X)$ can be found directly as a function of $\bar{S}(X)$, then ${\rho }_{\mathsf{s}\mathsf{e}}$ will be of the form ${\rho }_{\mathsf{s}\mathsf{e}}={p}_{0}|00\rangle \langle 00|+{p}_{1}|11\rangle \langle 11|$, which does not allow us to extract any EPR pairs, or equivalently, does not contain any secret information. Consequently, $\bar{S}$ and $\bar{E}$ should be as statistically independent as possible of $S$ and $E$.

We define $\bar{S}$ and $\bar{E}$ as the following cumulative probability functions: $\bar{S}(x)=\operatorname{Pr}[X\le x\: |\: S(X)=S(x)]$ and $\bar{E}({x}^{\prime },\bar{s},s)=\operatorname{Pr}[{X}^{\prime }\le {x}^{\prime }\: |\: \bar{S}(X)=\bar{s},S(X)=s,E({X}^{\prime },\bar{s})=E({x}^{\prime },\bar{s})]$. By definition, these functions are uniformly distributed between $0$ and $1$, independently of the other variables available to the party calculating it (Alice for $\bar{S}$ and Bob for $\bar{E}$). These functions also enjoy the property of making the subsystem ${\rho }_{\mathsf{s}\mathsf{e}}$ pure in absence of eavesdropping (i.e., when $\rho $ is pure), indicating that this choice of $\bar{S}$ and $\bar{E}$ does not introduce more impurity in ${\rho }_{\mathsf{s}\mathsf{e}}$ than $\rho $ already has.

For a pure state $|\psi \rangle =\int \mathrm{d}x\mathrm{d}{x}^{\prime }f(x,{x}^{\prime }){|x\rangle }_{{\mathsf{a}}_{1}}{|{x}^{\prime }\rangle }_{\mathsf{b}}$, with ${|x\rangle }_{{\mathsf{a}}_{1}}$ (resp. ${|{x}^{\prime }\rangle }_{\mathsf{b}}$) an $\mathbf{x}$-eigenstate in ${\mathsf{a}}_{1}$ (resp. in $\mathsf{b}$), the application of $\mathcal{QS}$ and $\mathcal{QE}$ gives: \begin{equation}\sum _{s,e\in \{0,1\}}\int \mathrm{d}\bar{s}\mathrm{d}\bar{e}\: \sigma (x)\epsilon ({x}^{\prime },\bar{s},s)f(x,{x}^{\prime }){|s\rangle }_{\mathsf{s}}{|\bar{s}\rangle }_{\bar{\mathsf{s}}}{|e\rangle }_{\mathsf{e}}{|\bar{e}\rangle }_{\bar{\mathsf{e}}}\text{,}\end{equation} where $x$ and ${x}^{\prime }$ are shorthand notations for $x(s,\bar{s})$ and ${x}^{\prime }(e,\bar{e},\bar{s})$ respectively. Let ${f}_{1}$ and ${f}_{2}$ be real and non-negative functions verifying $f(x,{x}^{\prime })={f}_{1}(x){f}_{2}(x,{x}^{\prime })$. ${f}_{1}(x)$ is chosen such that ${|{f}_{1}(x)|}^{2}$ is the probability density function of Alice's modulation, and ${f}_{2}(x,{x}^{\prime })$ such that ${|{f}_{2}(x,{x}^{\prime })|}^{2}$ is the probability density function of Bob's measured value ${x}^{\prime }$ conditionally to Alice sending $x$. Then, it is easy to check that we can factor ${\sum }_{a,b\in \{0,1\}}{\alpha }_{ab}{|ab\rangle }_{\mathsf{s}\mathsf{e}}$ out of $|\psi \rangle $ by setting:  \begin{align}
\sigma (x(s,\bar{s}))= & {\sigma }_{0}(s){({f}_{1}(x(s,\bar{s})))}^{-1}\text{,} \\
\epsilon (\bar{s},{x}^{\prime }(e,\bar{e},\bar{s}),s)= & {\epsilon }_{0}(e,s){({f}_{2}(x(s,\bar{s}),{x}^{\prime }(e,\bar{e},\bar{s})))}^{-1}\text{,}
\end{align}
 where \begin{align}
{\sigma }_{0}^{2}(s)= & {\int }_{x:S(x)=s}{|{f}_{1}(x)|}^{2}\mathrm{d}x\text{,} \\
{\epsilon }_{0}^{2}(e,s)= & {\int }_{x,{x}^{\prime }:S(x)=s,E({x}^{\prime },\bar{S}(x))=e}{|{f}_{2}(x,{x}^{\prime })|}_{2}\mathrm{d}x\mathrm{d}{x}^{\prime }\text{.}
\end{align}
 The conclusion follows from the definition of $\sigma $ and $\epsilon $.

 When more than one slice is involved, the functions $\bar{S}$ and $\bar{E}$ are defined similarly:

\begin{gather}\bar{S}(x)=\operatorname{Pr}[X\le x|{S}_{1\dots m}(X)={S}_{1\dots m}(x)]\text{,} \\
\begin{split}\bar{E}({x}^{\prime },\bar{s},{s}_{1\dots m})=\operatorname{Pr}[{X}^{\prime }\le {x}^{\prime }|\bar{S}(X)=\bar{s} \\
\wedge \: {S}_{1\dots m}(X)={s}_{1\dots m} \\
\wedge \: {E}_{1}({X}^{\prime },\bar{s})={E}_{1}({x}^{\prime },\bar{s})\wedge \dots \\
\wedge \: {E}_{m}({X}^{\prime },\bar{s},{s}_{1\dots m-1})={E}_{m}({x}^{\prime },\bar{s},{s}_{1\dots m-1})]\text{.}\end{split}\end{gather}\section{The Attenuation Channel}
\label{sectionAttenuationChannel}
    We now apply the slicing construction and display some results on the rates one can achieve in an important practical case. These results serve as an example and do not imply an upper bound on the achievable rates or distances. Instead, they can be viewed as lower bounds on an achievable secure rate in the particular case of an attenuation channel with given losses. Stated otherwise, this section simulates the rates we would obtain in a real experiment where Alice and Bob would be connected by an attenuation channel. For more general properties of the construction, please refer to Sec.~\ref{sectionAsymptotic}.

The purpose of this section is twofold. First, we wish to illustrate the idea of the previous section and show that it serves realistic practical purposes. Beyond the generality of the sliced error correction, its implementation may be easier than it first appears. Furthermore, the purification (resp. distillation) of more than one qubit (resp. bit) per sample is useful, as illustrated below.

Second, it is important to show that the construction works in a case as important as the attenuation channel. Clearly, requesting that a QKD protocol yields a non-zero secret key rate under all circumstances is unrealisitic---an eavesdropper can always block the entire communication. On the other hand, a QKD protocol that would always tell Alice and Bob that zero secure bits are available would be perfectly secure but obviously also completely useless. Of course, between these two extreme situations, the practical efficiency of a QKD protocol is thus important to consider.

The attenuation channel can be modeled as if Eve installed a beam-splitter in between two sections of a lossless line, sending vacuum at the second input. We here assume that Alice sends coherent states with a modulation variance of $31{N}_{0}$, with ${N}_{0}$ the vacuum fluctuations, which gives Alice and Bob up to $I(A;B)=2.5$ common bits in absence of losses or noise. This matches the order of magnitude implemented in \cite{bib:entry006}. We define the slices ${S}_{1}$ and ${S}_{2}$ by dividing the real axis into four equiprobable intervals labeled by two bits, with ${S}_{1}$ representing the least significant bit and ${S}_{2}$ the most significant one. More precisely, ${S}_{1}(x)=0$ when $x\le -\tau $ or $0<x\le \tau $ and ${S}_{1}(x)=1$ otherwise, with $\tau =\sqrt{2×31{N}_{0}}\: {\operatorname{erf}}^{-1}(1/2)$, and ${S}_{2}(x)=0$ when $x\le 0$ and ${S}_{2}(x)=1$ otherwise.

In this constructed example, we wish to calculate the theoretical secret key rate we would obtain in an identical setting. For various loss values, the secret key rates are evaluated by numerically calculating $\operatorname{Tr}(({\mathbf{Z}}_{{\mathsf{s}}_{i}}\otimes {\mathbf{Z}}_{{\mathsf{e}}_{i}})\rho )$, to obtain the bit error rates of slices $i=1,2$ and $\operatorname{Tr}(({\mathbf{X}}_{{\mathsf{s}}_{i}}\otimes {\mathbf{X}}_{{\mathsf{e}}_{i}})\rho )$ to obtain the phase error rates. Then, assuming asymptotically efficient binary codes, the rate is $R={R}_{1}+{R}_{2}={\sum }_{i=1,2}(1-h({e}_{i}^{\text{b}})-h({e}_{i}^{\text{p}}))$.

\begin{figure}
\begin{center}
\begin{tabular}{|r|r|r|r|r|r|r|}
\hline
& \multicolumn{3}{|c|}{$\rho_1$} & \multicolumn{3}{|c|}{$\rho_2$} \\
\cline{2-7}
Losses & $e_1^\text{b}$ & $e_1^\text{p}$ & $R_1$ & $e_2^\text{b}$ & $e_2^\text{p}$ & $R_2$ \\
\hline
0.0dB & 3.11\% & 5.33\% & 0.752 & 0.0000401 & 0.710\% & 0.938 \\
0.4dB & 3.77\% & 13.7\% & 0.193 & 0.0000782 & 28.6\% & 0.135 \\
0.7dB & 4.32\% & 20.0\% & 0.0204 & 0.000125 & 37.5\% & 0.0434 \\
\cline{2-4}
1.0dB & \multicolumn{3}{|c|}{-} & 0.000194 & 42.3\% & 0.0147 \\
1.4dB & \multicolumn{3}{|c|}{-} & 0.000335 & 45.6\% & 0.00114 \\
\hline
\end{tabular}
\end{center}

\caption{Error and EPR rates with two slices in an attenuation channel}
\label{tableAttenuationChannel}
\end{figure}
Using this two-slice construction, we were able to get the EPR rates described in Fig.~\ref{tableAttenuationChannel}. For the case with no losses, it is thus possible to distill $R=0.752+0.938=1.69$ EPR pairs per sample. Also, note that the phase error rate increases faster with the attenuation for ${\rho }_{2}$ than for ${\rho }_{1}$, with ${\rho }_{i}={\rho }_{{\mathsf{s}}_{i}{\mathsf{e}}_{i}}={\operatorname{Tr}}_{\text{All}\setminus \{{\mathsf{s}}_{i},{\mathsf{e}}_{i}\}}(\rho )$. This intuitively follows from the fact that the information Eve can gain from her output of the beam splitter affects first the most significant bit contained in ${S}_{2}(X)$.

 Due to the higher bit error rate in ${\rho }_{1}$, it was not possible to distill EPR pairs in slice~1 with losses beyond 0.7~dB. It was however still possible to distill EPR pairs in slice~2, up to 1.4~dB losses (about 10km with fiber optics with losses of 0.15dB/km). This result does not pose any fundamental limit, as it can vary with the modulation variance and with the choice of the functions ${S}_{1}$ and ${S}_{2}$. Note that the slice functions could be optimized in various ways, one of which being to use other intervals (as done in \cite{bib:entry010}, not necessary equiprobable and possibly chosen as a function of the losses), and another being to consider multi-dimensional slices as explained in the next section.

 Finally, note that although this example involves a Gaussian channel, this particularity is not exploited here and such a calculation can be as easily done for a non-Gaussian attack.

\section{Asymptotic Behavior}
\label{sectionAsymptotic}
    In this section, we study the behavior of the slice construction when the slice and slice estimator mappings take as input a block of $w$ states, with $w$ arbitrarily large. In \cite{bib:entry010}, the classical sliced error correction is shown to reduce to Slepian-Wolf coding \cite{bib:entry015} (asymmetric case with side-information) when using asymptotically large block sizes. We here study the quantum case, which is different at least by the fact that privacy amplification is explicitly taken into account.

For simplicity of the notation, we will study the asymptotic behavior in the case of an individually-probed channel only (although Eve's measurement can be collective). A study of periodic probing with period much smaller than the key size would give the same results, since in both cases it allows us to consider a sequence of identical random experiments and to study the typical case. However, joint attacks, with width as large as the key size, are outside the scope of this section, as the statistical tools presented here would not be suitable.

It is important to stress out that we here investigate what the secret key rates would be if the actual channel is an individually-probed one. The use of the protocol of this paper still requires to evaluate the phase error rate in all cases and this quantity is sufficient to determine the number of secret key bits. In the case of joint attacks, the secret key rates stated in the special cases below would then differ from the one obtained using the phase error rate.

\subsection{Direct reconciliation}
We thus here consider a block of $w$ states and the functions $S$, $\bar{S}$, $E$ and $\bar{E}$ on blocks of $w$ variables as well. Among the qubits produced by $\mathcal{QS}$, there is a certain number of them whose disclosed value allows Alice and Bob to correct (almost) all bit errors for the remaining slices. Then, among the remaining slices, a certain number of qubits allows Alice and Bob to correct (almost) all phase errors for the rest of the qubits. These last qubits are thus equivalent to secret key bits in the prepare-and-measure protocol.

We consider the following state, with the action of the channel modeled as joining system $\mathsf{b}$ with that of an eavesdropper Eve, and with $p$ left out as a public classical parameter: \begin{equation}
\label{eq_psi_ABE}|\Psi (p)\rangle =\int \mathrm{d}x\: g(x){|x\rangle }_{{\mathsf{a}}_{1}}{|\phi (x,p)\rangle }_{\mathsf{b},\mathsf{eve}}\text{.}\end{equation}  We consider $w$ such states coherently, and the mappings $\mathcal{QS}$ and $\mathcal{QE}$ take all $w$ states as input. We will follow the lines of the reasoning in \cite{bib:entry013, bib:entry016, bib:entry017} to show that the secret key rate tends to $I(X;{X}^{\prime })-I(X;E)$ for $w\to \infty $, with $X$ the random variable representing Alice's measure of ${\mathsf{a}}_{1}$ with $\mathbf{x}$, ${X}^{\prime }$ the measure of $\mathsf{b}$ with $\mathbf{x}$, and $I(X;E)=H(X)+H({\rho }_{\mathsf{eve}})-H({\rho }_{{\mathsf{a}}_{1},\mathsf{eve}})$, where $H(\rho )$ is the Von Neumann entropy of a state $\rho $.  The remainder of the discussion must be understood for any $\epsilon ,{\epsilon }_{U}>0$, with $w$ sufficiently large.

 Consider a mapping $U$ from $\mathbf{R}$ to a finite set $\mathcal{U}$ of size ${2}^{m}$, for some sufficiently large $m$, such that $I(U(X);{X}^{\prime })\ge I(X;{X}^{\prime })-{\epsilon }_{U}$. Let $\bar{S}(X)$ be the remaining continuous information not contained in $U(X)$, defined as in Sec.~\ref{sectionConstructionSbarAndEbar}. Let $x(\bar{s},u)$ be the mapping that recovers $x$ from $\bar{S}(x)$ and $U(x)$.

We here recall some definitions from \cite{bib:entry016}. For a given value of ${\bar{s}}^{(w)}$ and ${p}^{(w)}$ (${\bar{s}}^{(w)},{p}^{(w)}\in {\mathbf{R}}^{w}$), a HSW code \cite{bib:entry013, bib:entry017} $\mathcal{B}$ is a subset of ${\mathcal{U}}^{w}$ such that the corresponding $w$-wide states ${|{\phi }^{(w)}({x}^{(w)}({\bar{s}}^{(w)},{u}^{(w)}),{p}^{(w)})}_{{\mathsf{b}}^{(w)},\mathsf{eve}}$, ${u}^{(w)}\in \mathcal{B}$, can be distinguished by Bob with probability at least $1-\epsilon $. A privacy amplification (PA) set $\mathcal{E}$ is a subset of ${\mathcal{U}}^{w}$ such that the sum of the corresponding states ${\sum }_{{u}^{(w)}\in \mathcal{E}}{|{\phi }^{(w)}({x}^{(w)}({\bar{s}}^{(w)},{u}^{(w)}),{p}^{(w)})\rangle }_{{\mathsf{b}}^{(w)},\mathsf{eve}}$ factors Eve. Finally, a key generation code $\mathcal{B}$ is a HSW code that can be divided into a collection of non-overlapping PA sets $\mathcal{B}={\cup }_{k}{\mathcal{E}}_{k}$. In the sequel, we drop the $w$ superscript for simplicity.

 Consider three consecutive ranges $I=1\dots |I|$, $J=|I|+1\dots |I|+|J|$, $K=|I|+|J|+1\dots |I|+|J|+|K|$ with sizes $|I|=\lceil wH(U(X)|{X}^{\prime })+\epsilon \rceil$, $|J|=\lceil wI(U(X);E)\rceil$ and $|K|=\lfloor wI(U(X);{X}^{\prime })-wI(U(X);E)-\epsilon \rfloor$. Note that $|I|+|J|+|K|\le wH(U(X))+2\le wm+2$. These three ranges will correspond to three kind of slices in the derived prepare-and-measure protocol: ${S}_{I}$, ${S}_{J}$ and ${S}_{K}$. ${S}_{I}$ will give Bob enough information to perform error correction, ${S}_{J}$ will contain bits, equal between Alice and Bob, which will be sacrificed with PA since they are not necessarily secret, and ${S}_{K}$ will contain equal and secret bits (i.e., key bits).

 From \cite{bib:entry016}, it is possible to cover the space of $wm+2$-bit vectors with ${2}^{|I|}$ key generation codes ${\mathcal{C}}_{{s}_{I}}$ of size ${2}^{|J|+|K|}$. To each element of ${\mathcal{U}}^{w}$, we assign a $|I|$-bit vector that identifies the key generation code it belongs to; this defines the first $|I|$ slices ${S}_{I}(X)$.

 By providing the bit syndromes ${\xi }_{I}^{\text{b}}$ to Bob, he can identify ${S}_{I}(X)={s}_{I}$ and the associated key generation code ${\mathcal{C}}_{{s}_{I}}$. By definition, he has enough information to identify an element within it. Such an element can be uniquely labeled by a $|J|+|K|$-bit vector, thereby defining ${S}_{J}$ and ${S}_{K}$. So, there exists a mapping that maps ${|{s}_{I}\rangle }_{{\mathsf{s}}_{I}^{\prime }}{|\phi (x(\bar{s},u),p)\rangle }_{\mathsf{b},\mathsf{eve}}$ onto ${|{s}_{I}\rangle }_{{\mathsf{s}}_{I}^{\prime }}{|{s}_{JK}\rangle }_{{\mathsf{e}}_{JK}}{|{\phi }^{\prime }(x(\bar{s},u),p)\rangle }_{\bar{\mathsf{e}},{\mathsf{e}}_{{I}_{,}}\mathsf{eve}}$ with probability at least $1-\epsilon $, and thus ${e}_{i}^{\text{b}}\le \epsilon $, $\forall i\in J\cup K$.

 Each key generation code contains ${2}^{|K|}$ PA sets of size ${2}^{|J|}$ each \cite{bib:entry016}. The labeling can be such that ${S}_{K}$ corresponds to the identification of the PA set and ${S}_{J}$ the element inside the PA set.

 Providing the phase syndromes ${\xi }_{J}^{\text{p}}$ to Bob gives him enough information to determine the phase of Alice's qubits in ${\mathsf{s}}_{J}$. If the phase errors are corrected by Bob, measuring or tracing out subsystems ${\mathsf{s}}_{J}{\mathsf{e}}_{J}$ is equivalent to summing the slices in $J$ over all possible bit values and thus factoring out Eve. More precisely, with $\bar{s}$, ${s}_{I}$ and $p$ fixed (and the corresponding subsystems not shown), and with ${|{s}_{j}^{*}\rangle }_{{\mathsf{s}}_{j}^{*}}={2}^{-1/2}({|0\rangle }_{{\mathsf{s}}_{j}}+{(-1)}^{{s}_{j}^{*}}{|1\rangle }_{{\mathsf{s}}_{j}})$ (and similarly for ${\mathsf{e}}_{j}$ and ${\mathsf{s}}_{j}^{\prime }$), the system after correction of ${s}_{I}$ is of the form: \begin{multline}|\Psi \rangle =\sum _{{s}_{K}{s}_{J}}{|{s}_{JK}\rangle }_{{\mathsf{s}}_{JK}}{|0\rangle }_{{\mathsf{s}}_{J}^{\prime }}{|{s}_{JK}\rangle }_{{\mathsf{e}}_{JK}}{|{\phi }^{\prime }({s}_{JK})\rangle }_{\bar{\mathsf{e}},{\mathsf{e}}_{I},\mathsf{eve}}\\
=\sum _{{s}_{K}{s}_{J}{s}_{J}^{*}{e}_{J}^{*}}{(-1)}^{{s}_{J}({s}_{J}^{*}+{e}_{J}^{*})}{|{s}_{J}^{*}\rangle }_{{\mathsf{s}}_{J}^{*}}{|{s}_{K}\rangle }_{{\mathsf{s}}_{K}}\\
\otimes {|0\rangle }_{{\mathsf{s}}_{J}^{\prime }}{|{e}_{J}^{*}\rangle }_{{\mathsf{e}}_{J}^{*}}{|{s}_{K}\rangle }_{{\mathsf{e}}_{K}}{|{\phi }^{\prime }({s}_{JK})\rangle }_{\bar{\mathsf{e}},{\mathsf{e}}_{I},\mathsf{eve}}\text{.}\end{multline} Then Alice sends to Bob information about her phase (${s}_{J}^{*}$), which he stores in his auxiliary register ${\mathsf{s}}_{J}^{\prime }$. The state becomes: \begin{multline}\sum _{{s}_{K}{s}_{J}{s}_{J}^{*}{e}_{J}^{*}}{(-1)}^{{s}_{J}({s}_{J}^{*}+{e}_{J}^{*})}{|{s}_{J}^{*}\rangle }_{{\mathsf{s}}_{J}^{*}}{|{s}_{K}\rangle }_{{\mathsf{s}}_{K}}\\
\otimes {|{s}_{J}^{*}\rangle }_{{\mathsf{s}}_{J}^{\prime }}{|{e}_{J}^{*}\rangle }_{{\mathsf{e}}_{J}^{*}}{|{s}_{K}\rangle }_{{\mathsf{e}}_{K}}{|{\phi }^{\prime }({s}_{JK})\rangle }_{\bar{\mathsf{e}},{\mathsf{e}}_{I},\mathsf{eve}}\text{.}\end{multline} The difference between Alice's and Bob's phases is calculated in ${\mathsf{s}}_{J}^{\prime }$ and the correction is applied to ${\mathsf{e}}_{J}^{*}$. Overall, this transformation can be summarized as ${|{s}_{J}^{*}\rangle }_{{\mathsf{s}}_{J}^{\prime }}{|{e}_{J}^{*}\rangle }_{{\mathsf{e}}_{J}^{*}}\to {|{s}_{J}^{*}+{e}_{J}^{*}\rangle }_{{\mathsf{s}}_{J}^{\prime }}{|{s}_{J}^{*}\rangle }_{{\mathsf{e}}_{J}^{*}}$. This gives the following state: \begin{multline}\sum _{{s}_{K}{s}_{J}^{*}}{|{s}_{J}^{*}\rangle }_{{\mathsf{s}}_{J}^{*}}{|{s}_{K}\rangle }_{{\mathsf{s}}_{K}}{|{s}_{J}^{*}\rangle }_{{\mathsf{e}}_{J}^{*}}{|{s}_{K}\rangle }_{{\mathsf{e}}_{K}}\\
\otimes \sum _{{s}_{J},({s}_{J}^{*}+{e}_{J}^{*})}{(-1)}^{{s}_{J}({s}_{J}^{*}+{e}_{J}^{*})}{|{s}_{J}^{*}+{e}_{J}^{*}\rangle }_{{\mathsf{s}}_{J}^{\prime }}{|{\phi }^{\prime }({s}_{JK})\rangle }_{\bar{\mathsf{e}},{\mathsf{e}}_{I},\mathsf{eve}}\\
=\sum _{{s}_{K}{s}_{J}^{*}}{|{s}_{J}^{*}\rangle }_{{\mathsf{s}}_{J}^{*}}{|{s}_{K}\rangle }_{{\mathsf{s}}_{K}}{|{s}_{J}^{*}\rangle }_{{\mathsf{e}}_{J}^{*}}{|{s}_{K}\rangle }_{{\mathsf{e}}_{K}}\\
\otimes \sum _{{s}_{J}}{|{s}_{J}\rangle }_{{\mathsf{s}}_{J}^{\prime *}}{|{\phi }^{\prime }({s}_{JK})\rangle }_{\bar{\mathsf{e}},{\mathsf{e}}_{I},\mathsf{eve}}\text{.}\end{multline} Finally, the sum $\sum _{{s}_{J}}{|{s}_{J}\rangle }_{{\mathsf{s}}_{J}^{\prime *}}{|{\phi }^{\prime }({s}_{JK})\rangle }_{\bar{\mathsf{e}},{\mathsf{e}}_{I},\mathsf{eve}}$ factors out Eve, by definition of a PA set.

Given the size of $K$, we conclude that the secret bit rate can asymptotically come as close as desired to $I(X;{X}^{\prime })-I(X;E)$. Note that in the particular case of the attenuation channel, an evaluation of the secret key rate can be found in \cite{bib:entry008, bib:entry009}.

\subsection{Reverse reconciliation}
So far, we have always assumed that the slices apply to Alice and the slice estimators to Bob. However, there are some cases for which the opposite case increases the secret bit rate \cite{bib:entry006}.

Let us start again from the state $|\Psi (p)\rangle $ as in (\ref{eq_psi_ABE}), and rewrite ${|\phi (x,p)\rangle }_{\mathsf{b},\mathsf{eve}}$ as ${|\phi (x,p)\rangle }_{\mathsf{b},\mathsf{eve}}=\int \mathrm{d}{x}^{\prime }\: f(x,p,{x}^{\prime }){|{x}^{\prime }\rangle }_{\mathsf{b}}{|\phi (x,p,{x}^{\prime })\rangle }_{\mathsf{eve}}$. Let $h({x}^{\prime },p)$ be a non-negative real function such that ${h}^{2}({x}^{\prime },p)=\int \mathrm{d}x{|g(x,p)f(x,p,{x}^{\prime })|}^{2}$. Then, \begin{multline}|\Psi (p)\rangle =\int \mathrm{d}{x}^{\prime }\: h({x}^{\prime },p){|{x}^{\prime }\rangle }_{\mathsf{b}}{|{\phi }^{\prime }({x}^{\prime },p)\rangle }_{{\mathsf{a}}_{1},\mathsf{eve}}\text{,}\\
\text{with }{|{\phi }^{\prime }({x}^{\prime },p)\rangle }_{{\mathsf{a}}_{1},\mathsf{eve}}=\\
\int \mathrm{d}x\: g(x,p)f(x,p,{x}^{\prime })/h({x}^{\prime },p){|x\rangle }_{{\mathsf{a}}_{1}}{|\phi (x,p,{x}^{\prime })\rangle }_{\mathsf{eve}}\text{.}\end{multline} Thus, by applying the same argument as for direct reconciliation, we can asymptotically reach $I(X;{X}^{\prime })-I({X}^{\prime };E)$ secret bits when $\mathcal{Q}\mathcal{S}$ is applied on system $\mathsf{b}$ and $\mathcal{Q}\mathcal{E}$ on system ${\mathsf{a}}_{1}$. The evaluation of the secret key rate for reverse reconciliation can also be found in \cite{bib:entry008, bib:entry009}.

\section{Conclusion}
\label{sectionConclusion}
    In this paper, we studied the equivalence between an EP protocol and a QKD protocol with sliced error correction for reconciliation. In the QKD protocol, Alice sends Gaussian-modulated coherent states to Bob, who measures the result using homodyne detection. To probe the channel and determine the amount of entanglement that can be transmitted through it, Bob has to make homodyne measurements in all quadratures.

We found that the EP protocol based on sliced error correction is indeed efficient and allows its equivalent prepare-and-measure QKD protocol to produce a secret key which is secure against any eavesdropping strategy. Although the qubit encoding scheme is derived from a reconciliation protocol easily implementable in practice \cite{bib:entry006}, the main drawback of the method is the possibly huge number of measurements to get a statistically relevant estimation of the phase error rate and thus the number of secret key bits. Yet in theory, the sample set can be reduced to an arbitrarily small fraction of the produced key, when an arbitrarily large number of quantum states are processed through secret key distillation.

An advantage of this method is that it can in principle be adapted to other modulation distributions---the fact that the modulation is Gaussian is not crucial. In practice, the finite range of the amplitude modulator does not allow one to produce a real Gaussian distribution for the prepare-and-measure protocol, and one can take this effect explicitly into account. Also, it may be more efficient to consider a modulation of coherent states along a uniform distribution over a finite domain of $(x,p)$ (e.g., a square or a circle centered on $(0,0)$) so as to increase the correlations between Alice and Bob.

Open problems for further research include the improvement of the statistical estimation of the EP parameters, the investigation of other modulation distributions and the optimization of the encoding scheme for practical implementations.

\section*{Acknowledgments}
This work was presented at the ESF Continuous Variables Quantum Information Processing Workshop 2004, Veilbronn, Germany, April 2004 and in part at the ESF Continuous Variables Quantum Information Processing Workshop 2003, Aix-en-Provence, France, April 2003. We acknowledge financial support from the Communauté Française de Belgique under grant ARC 00/05-251, from the IUAP programme of the Belgian government under grant V-18 and from the EU under project RESQ (IST-2001-35759). S.I. acknowledges support from the Belgian FRIA foundation, as well as the Swiss NCCR and the European Project RESQ. We acknowledge discussions with Frédéric Bourgeois, Jaromír Fiurá\v sek, Philippe Grangier, Frédéric Grosshans, Patrick Navez, John Preskill and Serge Van Criekingen.

\end{document}